# Network Anomaly Detection: Flow-Based or Packet-Based Approach?

Huy Anh Nguyen, Deokjai Choi
Department of Computer Engineering, Chonnam National University
Email: anhhuy@gmail.com, dchoi@chonnam.ac.kr

ABSTRACT

One of the most critical tasks for network administrator is to ensure system uptime and availability. For the network security, anomaly detection systems, along with firewalls and intrusion prevention systems are the must-have tools. So far in the field of network anomaly detection, people are working on two different approaches. One is flow-based; usually rely on network elements to make so-called flow information available for analysis. The second approach is packet-based; which directly analyzes the data packet information for the detection of anomalies. This paper describes the main differences between the two approaches through an in-depth analysis. We try to answer the question of when and why an approach is better than the other. The answer is critical for network administrators to make their choices in deploying a defending system, securing the network and ensuring business continuity.

## I. INTRODUCTION

Operators of mission critical networks employ a variety of strategies to ensure system uptime and availability. To secure the network from outside malicious activities, firewalls and intrusion prevention systems (IPSs) maybe utilized, along with performance measurement tools and network infrastructure health monitoring systems. However, to protect networks against threads such as DDoS attacks and worm outbreaks, intelligent, real-time solutions are needed.

Such kinds of anomalies generate vast amounts of bogus traffic, which can overwhelm the network and any attached hosts. In addition, the traffic that is generated by anomalies may not have a signature, which is required by a typical IPS. It may also arrive on otherwise completely legitimate ports, passing the security checks of a firewall. As a result, a new category of network security systems has appeared, specifically geared to solve this problem. These systems utilize what is commonly known as *Behavioral Anomaly Detection* or *Network Behavior Analysis*. Rather than just looking at volumes of packets, these systems intelligently take into account the behavior of the network and the hosts that are attached to that network. Changes in the network behavior are used to detect DDoS attacks, worm outbreaks and otherwise misbehaving hosts or network elements with dramatically improved accuracy. As more and more administrators of mission critical networks recognize that an additional layer of security is needed besides the traditional signature based systems (i.e. IPSs and firewalls ...), it has become best-practice to deploy an intelligent behavioral anomaly detection solution in the networks, along with the already existing security infrastructure.

This paper describes the main differences between these two approaches by analyzing the most important features regarding security of the network. During the analysis, we also made discussions on common beliefs about the two approaches. Due to the existence of strong biases in people's opinions, discussions are needed to have a clear and fair review.

## Ⅱ. FLOW-BASED ANOMALY DETECTION

Flow-based anomaly detection centers around the concept of the network flow. A flow record is a summarized indicator that a certain network flow took place and that two network end points have communicated with each other at some time in the past. A flow record typically contains the IP network addresses of the two hosts, network ports, network protocol, amount of data that was sent as part of this connection, the time when the flow occurred as well as a few miscellaneous flags.

Flow-based approaches rely heavily on the ability of network devices to generate flow information. A typical anomaly detection system using flow information would be implemented as depicted in Fig. 1. An anomaly detection component would sit right behind the router to collect all flows going in / out of the network for analyzing. There are many solutions from different vendors to generate such information. For example Cisco[1] and Juniper[2] routers are capable of observing network traffic and generating NetFlow data. Or solutions from Foundry,

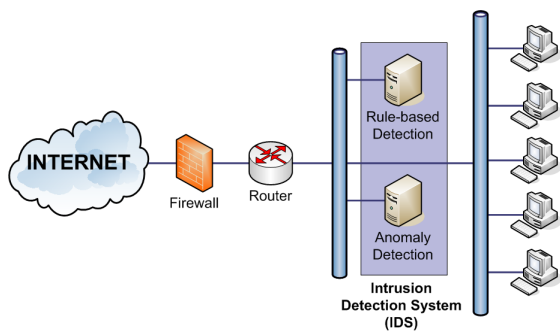

Figure 1. Anomaly Detection System Implementation Map.

Extreme and HP ProCurve allow us to have flow information with similar structure to NetFlow, which is called sFlow[3]. There also be solutions from open-source projects that generate nFlow data. These flow records are then written into newly created packets and sent off to a recipient (usually through UDP protocol) for analysis.

Flow records are well suited to represent the interactions between hosts in a network. By analyzing exported flow records and looking for unusual amounts, directions, groupings and characteristics of flows, an anomaly detection solution can infer the presence of worms or DDoS attacks in a network. Many solutions for flow-based anomaly detection from different vendors are available, among which, Lancope[4] and Arbor Networks provide the currently best-value security systems on the market. They both utilize a mixture of detection methodologies that include both pure anomaly detection as well as algorithmic pattern matching.

## Ⅲ. PACKET-BASED ANOMALY DETECTION

A packet-based anomaly detection system can also be implemented as in Fig. 1, but unlike flow-based solutions, does not rely on third-party components to generate meta or summary information of the network traffic. Instead, all analysis is based on observed raw packets, as they traverse the network links and captured by network devices.

There are several methods in which the network traffic can be captured for analyzing. One is to configure a spanning port. A router or switch then makes a copy of every packet that is sent / received on one or more of its interface ports, and sends copy out of the span port. Another method, which is more preferable is more than one ways, is the use of network taps[6]. Those are passive devices, which allow the fully transparent observation of packets on a network link.

Once a packet-based anomaly detection solution is set up, statistics about the observed packets are accumulated and analyzed by a variety of methods. For example, Esphion's netDeFlect and CounterStorm's UPAD uses sophisticated neural networks to detect the presence of anomalous traffic. In addition, the content within those packets is kept and can be used for further advance anomaly detection.

## IV. FLOW-BASED VS. PACKET-BASED

When comparing these two methods of anomaly detection, the architecture plays and important role[5]. Some networks lend themselves more to one approach than to the other. Another important factor when choosing between the two approaches is the experience and personal reference of the network administrator. Due to these natural biases, we would like to make a fair comparison between the two approaches based on main features regarding security of the network.

1. Network Scale

As stated above, the architecture plays the major role in which administrator will choose what kind of approach to be implemented. Small and medium enterprise networks would like to implement packet-based approaches since the incoming / outgoing traffic of their networks is manageable. In such cases, network administrators of these networks would like to have the ability to closely and directly manage traffic data. Packet-based anomaly detection will be preferred due to its ease of use and simplification in deployment.

On the other hand, large networks, ISPs and even larger service provider would most likely show their interest in a flow-based approach, although they can deploy packet-based systems for its easy and useful network management. Traffic measurement will be more difficult in the next-generation Internet with features of high-speed links or new protocols such as IPv6 or MIPv6. In that case, flow-based approach with the ability to operate in very high bandwidth links (1, 5, or even 10 Gig+) is preferred. One other advantage of flow information in this case is that it mainly evaluated for accounting purpose, which is the main function of ISPs.

2. Deployment Cost

In small and medium size networks, packet-based approaches can be easily deployed using port spanning or deploying network taps. These devices functioning as traffic collection points, has the main purpose to collect network packets traveling through it and send these data to the data management center for analyzing. But the story will be different when we use packet-based approaches in large networks with fully meshed networks. Deploying probes throughout the network is an expensive task for both literal and figurative meaning. Even if money were no object, it would become a major effort to maintain a dozen or so probes over time and one would quickly find out that they are seldom located where they need them to analyze the data. To compound the issue, problems tend to be intermittent and disappear as quickly as they appeared. Besides, moving target analyzers to the strategic best physical location when problems arise is often geographically challenging in even medium networks.

Problems described above obviously do not exist in networks using flow-based approaches, which provide network operators the ability to create "virtual monitoring points" in the network. This reduces the total cost of ownership and deployment complexity. Because flow information enable visibility into many different points in the network at one time, they offer an uncanny ability to "connect the dots" between events as they make their way across the network from one geographic site to another. Thus contributing a huge leap forward in forensics analysis and auditing operations.

3. Data source

Probably one of the most important aspects of maintaining network security is the access to raw packet data for further in-depth analysis of network activities. Packet-based approaches, by its nature to capture all the packets, give users an excellent ability to store all the traffic data for real-time or further network investigation. Flow-based solutions on the contrary, only see summary records, produced by network devices, and therefore don't have access to raw data information, which is often vital for analysis and mitigation of an anomaly.

Another difference in the data source between the two approaches is the data size. Packet-based solutions tend to build fine-grained, high-volume packet traces. For example, an administrator may want to save all incoming TCP traffic for further investigation. In the case of high traffic network, storing for all these data may be very costly. The problem can be meliorated by techniques such as random data sampling, adaptive data sampling or partial data storing... In contrast, flow-level data only contain aggregated information which are coarse-grained and low-volume data[7]. In large networks, storing flow-data may also be an expensive problem, but much more affordable compare to packet-data solutions.

4. Low-latency Anomaly Detection

Routers and switches usually export a flow after there has been a certain time of inactivity, typically 5 to 15 seconds[8]. Thus, a flow-based solution can at the earliest only begin to detect an anomaly at least 5 - 15 seconds after its onset. In fact, network administrators can configure their infrastructures and set the flow export interval down to 1 second [1]. But in practical deployment, rarely do we see such a coarse-grained configured system. After flow information being exported, the detection algorithms can start to do their job, which may add some more time before actually coming to the conclusion that there is an anomaly.

On the contrary, a packet-based solution works in almost real-time. There is no 5 seconds lag before the statistical data about the network traffic is available. The detection algorithms continuously work on this real-time data. As a result, a packet-based solution can detect network anomalies faster than a flow-based solution.

One might argue that 5 seconds lag would be nothing, it would make no difference. But in some cases, it may be everything. Consider the case of an enterprise network. Coming back from a business trip, one of the employees plugs the laptop back in. Unfortunately, while on travels, this laptop was infected by an aggressively scanning worm, which now starts to look for new victims in the company's network. Depending on the exact scanning algorithm of the worm, an infection of another machine may happen within seconds. Therefore, every second counts for the successful containment of the outbreak, and the 5 seconds lag may make all the difference.

5. Anomaly Source Trace

Now we consider a co-operation scenario between two companies. A network professional at Company A receives a phone call from another Company B stating that someone in Company A is sending SNMP gets to the internet router at Company B which in turn was causing alerts to be sent via SNMP traps to the Network Management Station (NMS) at Company B.

With a packet-based anomaly detection system, administrators from Company A can start solving the problem by asking Company B for the IP address of its router. Loaded with the destination IP address, the network administrator from Company A bring out a laptop and make a visit to the data room in a different building. After booting the laptop, a telnet into the switch is performed and port spanning is configured so that the laptop sees all traffic to the Internet port. Then the administrator issues a query to filter IP that has communication with Company B's router. The malicious host is then identified and locked down.

However, with a flow-based solution available, the network administrator from Company A could have avoided packing up the laptop, walking to another building and switching up the port spanning. He simply has to search for the destination IP address and figured out who was communicating with Company B's router. In this case, flow information is much easier and faster in tracing down the anomaly source.

6. Miscellaneous

Still, there exist many controversies in whether an approach is superior to the other. One may argue that a flow-based detection solution relies on third-party network elements, such as routers and switches, to produce the flow-records that are its only insight into the current network traffic. And since not all the routers and switches are capable of producing flow information, flow-based solution is inflexible and can't be applied every where. Packet-based solution in this case, only rely on the ability to capture traffic packets from network interfaces, is much more preferable. However, at this moment, such a claiming is not true in most of the cases. Almost all the routers and switches from big network vendors are capable with the flow-producing ability. Examples are routers and switches from Cisco, Juniper, Foundry, Extreme and HP ProCurve... They all supported flow collecting for years and administrators simply need to turn it on. Therefore, flows, just like packets, for most companies are free and easy to use.

Another public belief states that flow-based solutions can not work accurately, especially under heavy load. The reason for that claiming is that flow-based solutions place an overhead on network devices that can export flow information.

Under heavy work load, the problem may be severe. One solution that is often suggested is to use flow-sampling. The idea is not to consider every packet for the generation of flows, but only every $n^{th}$ packet, for example, every $100^{th}$ packet. Obviously, the number of generated flows is dramatically reduced, along with CPU load and network utilization. However, this comes at the price of lost accuracy. Any information about the average flow length, average flow data or numbers of flows... will then become unreliable. In real-life scenarios, this problem does exist, but not so dramatic. An administrator may skillfully configure his network follow an adaptive flow sampling mechanism. For example, the network will increase the flow sampling as soon as there are some malicious activities spotted. This issue is currently an active research field with many solutions from researchers.

As flow-based and packet-based approaches show their shortcomings by one way or another, it's starting to come to existence of flow + packet based solutions. An example of such a mixed system is Lancope's StealthWatch solution, which not only can stop threats that are visible at the enterprise level using flow information but also allows for full packet capture and analysis[4]. Flow information will be assembled on a packet-by-packet basis within the anomaly detection system. Such solutions provide the ability to co-operate between flow and packet data, which compensate for each other.

## V. CONCLUSION

"There is no remedy for all cures" - as people usually say. So the same when we choose a solution between flow-based and packet-based anomaly detection. Both approaches show their strengths and weaknesses in particular conditions. In this paper, we made an in-depth comparison between the two approaches. We pointed out when and how an approach will be considered better than the other. We also argue on the common biases and explained the truth behind people's beliefs. The analyzing in this paper, though going into great details, is still based on our experiences of network management and other paper works. For the further research, we would like to conduct real experiments to statistically compare the performance between the two approaches. Only then we will have real deep understanding on how to choose the best solution for the network.

## REFERENCES

[1] Cisco IOS Software NetFlow
    http://www.cisco.com/warp/public/732/netflow
[2] Juniper Networks: JUNOS 7.2 Documentation
    http://www.juniper.net/techpubs/software/junos/junos72/index.html
[3] RFC 3176: sFlow Specification
    http://tools.ietf.org/html/rfc3176
[4] Lancope's Stealthwatch Flow Collector Apps
    http://lancope.com/downloads/StealthWatchFlowCollector.pdf
[5] Packet vs flow-based anomaly detection
    http://www.esphion.com/pdf/ESP_WP_4_PACKET_V_FLOWS.pdf
[6] Tap and Span Port Comparison
    http://www.netoptics.com/products/pdf/taps-and-span-ports.pdf
[7] Robin Sommer and Anja Feldman, "NetFlow: Information loss or win?", *Internet Measurement Workshop*, 2002.
[8] Adam Powers, "Face-off: Anomaly detecion", *NetworkWorld.com*, Mar. 2006.
[9] Jianning Mai, Chen-Nee Chuah, "Is Sampled Data Sufficient for Anomaly Detection?", *Proc. of IMC'06*, Oct. 2006.